\documentclass[
    ,final            % use final for the camera ready runs
]
 {aipproc}

\layoutstyle{6x9}

\begin{document}

\title{Weighted Configuration Model}


\author{M. \'Angeles Serrano}{
  address={Departament de F{\'\i}sica Fonamental, Universitat de
 Barcelona, Mart\'{\i} i Franqu\`es 1, 08028 Barcelona, Spain}
}

\author{Mari{\'a}n Bogu{\~n}{\'a}}
{
 address={Departament de F{\'\i}sica Fonamental, Universitat de
 Barcelona, Mart\'{\i} i Franqu\`es 1, 08028 Barcelona, Spain}
}

\begin{abstract}
 The configuration model is one of the most successful models for generating uncorrelated random networks. We analyze its behavior when the expected degree sequence follows a power law with exponent smaller than two. In this situation, the resulting network can be viewed as a weighted network with non trivial correlations between strength and degree. Our results are tested against large scale numerical simulations, finding excellent agreement.
 
 \end{abstract}

\maketitle

%%%%%%%%%%%%%%%%%%%%%%%%%%%%%%%%%%%%%%%%%%%%
%% MAINMATTER
%%%%%%%%%%%%%%%%%%%%%%%%%%%%%%%%%%%%%%%%%%%%

\section{Introduction}

Complex networked systems represent better than any other the idea of complexity. When looking at their large scale topological properties, real networks are far more complex than classical random graphs \cite{Erdos59,Erdos60}, showing emerging properties not obvious at the level of their elementary constituents --the small-world effect, scale-free connectivity, clustering, degree correlations, etc. When looking at the dynamical processes that take place on top of them, these large scale topological properties have striking consequences on the behavior of the system --absence of epidemic threshold, resilience to damage, etc. The understanding that many real world systems of interacting elements can be mapped into graphs or networks has led to a surge of interest in the field of complex networks and to the development of a theoretical frameworks able to properly analyze them \cite{Mendesbook,Barabasi02,Dorogovtsev02}. Under this approach, the elements of the system are mapped into vertices whereas interactions among these elements are represented as edges, or links, among vertices of the network.
 
 In an attempt to bring nearer theory and reality, many researchers
working on the new rapidly evolving science of complex
networks have recently shifted focus from unweighted graphs
to weighted networks\cite{Newman2001,Yook2001,Noh2002,Barrat04a,Barrat04b,Newman2004}. Commonly, interactions between
elements in network-representable complex real systems -may they
be communication systems, such as the Internet, or transportation
infrastructures, social communities, biological or biochemical
systems- are not of the same magnitude. It seems natural that the
first more simple representations, where edges between pairs of
vertices are quantified just as present or absent, give way to
more complex ones, where edges are no longer in binary states but
may stand for connections of different strength.

In an unweighted network, all the topological properties can be expressed as a function of the adjacency matrix, $A_{ij}$, whose elements take the value $1$ when vertices $i$ and $j$ are connected by an edge and $0$ otherwise. Using this matrix, the degree of a vertex --the number of neighbors or connections it has-- becomes
\begin{equation}
k_i=\sum_j A_{ij}.
\end{equation}
The distribution of $k_i$, $P(k)$, is called the degree distribution of the network, and it arises as the most fundamental topological characteristic. Surprisingly, in a vast majority of cases, real networks show degree distributions following power laws of the form $P(k) \sim k^{-\gamma}$ for $k \gg 1$ and $2<\gamma<3$. This implies that the second moment of the degree distribution, $\langle k^2 \rangle$, diverges in the thermodynamic limit, which causes the loss of any characteristic degree scale. For this reason, these class of networks are called scale-free (SF). This feature of the degree distribution is, eventually,  the responsible for the surprising behavior of dynamical processes that run on top of these networks.

For weighted networks, the adjacency matrix is no longer the fundamental quantity ruling the properties of the network. Instead, we have to consider the matrix $\omega_{ij}$, which measures the weight between the pair of vertices $i$ and $j$, that is, the magnitude of the connection between $i$ and $j$. The analogous quantity to the vertex degree is now the vertex strength, defined as
\begin{equation}
s_i=\sum_j \omega_{ij} A_{ij}.
\end{equation}
This quantity measures the total strength of vertex $i$ as the sum of the weights of all its connections. If the weights of a given vertex are not correlated with the vertex degree, the average strength of a vertex of degree $k_i$ is simply given by
\begin{equation}
\bar{s}(k_i)=\langle \omega \rangle k_i,
\end{equation}
where $\langle \omega \rangle$ is the average weight of the edges of the network. In this situation, strength and degree are proportional and weights do not incorporate more information to the network than that already present in the adjacency matrix. Real networks show, however, a very different scaling, with a non trivial relation between strength and degree of the form
\begin{equation}
s_i \sim k_i^{\beta},
\end{equation}
with $\beta \ne 1$ (typically $\beta >1$). This anomalous scaling reveals the presence of correlations between strength and degree and, thus, the need to model network formation in an weighted basis, where the evolution of the network is linked to the evolution of the weights assigned to connections. In this paper, we present a model of weighted network based on the configuration model. We shall show that, when the expected degree sequence follows a power law with exponent smaller than 2, the resulting network is weighted and shows a non trivial scaling between strength and degree.

\section{The configuration model}

The configuration model was first introduced as an algorithm to generate uncorrelated random networks with a given degree distribution, $P(k)$ \cite{Bekessy1972,Bender1978,Molloy1995,Molloy1998}. The model operates in two steps:
\begin{itemize}
\item
We start assigning to each vertex $i$, out of a set of $N$, a number of ``stubs'', $k_i$, drawn from the distribution $P(k)$, under the constraint that the sum $\sum_i k_i$ is even.
\item
Pairs of these stubs are chosen uniformly at random and the corresponding vertices are connected by an undirected edge.
\end{itemize}
Given the random nature of the edge assignment, this algorithm generates networks with the expected degree distribution and no correlations between the degrees of connected vertices. The model, as described above, allows the formation of multiple or self connections among vertices. Nevertheless, when the expected degree distribution has bounded fluctuations, the number of such ``pathological'' connections is small an can be neglected in the thermodynamic limit. In this case, one can add an extra constraint in step two of the algorithm avoiding multiple or self connections without modifying the resulting network. 

The picture changes drastically when the expected degree distribution has unbounded fluctuations. This is precisely the case of SF distributions with exponent $\gamma \in (2,3]$. In this situation, vertices with expected degree satisfying $k_i > \sqrt{\langle k \rangle N}$ cannot avoid to form multiple connections. Indeed, as it was proved in \cite{Boguna2004}, the number of multiple connections in this case scales as $N^{3-\gamma} \ln N$. Yet the situation is not so dramatic since the overall number of connections is much larger that the number of multiple connections. However, one has to be careful because the ``hubs'' of the network --those in the tail of the distribution-- are precisely the ones more prone to hold multiple connections and, therefore, it could alter the results of dynamical processes evaluated on top of these class of networks. There are two strategies one can follow in order to avoid multiple connections: 
\begin{itemize}
\item One can introduce a constraint in step two prohibiting multiple connections. This has the side effect of introducing strong disassortative degree correlations among connected vertices. 
\item Alternatively, one can introduce a cut-off in the distribution $P(k)$ scaling as $\sqrt{N}$. By doing so, one recovers uncorrelated networks but with a smaller cut-off than the natural one \cite{Catanzaro2004}.
\end{itemize}

\section{The configuration model beyond $\gamma=2$}

The configuration model with expected degree distribution with $\gamma \le 2$ has not been studied so forth. This is an extreme situation in which the number of multiple connections cannot be neglected. We can take advantage of this fact to construct a weighted network, where the weight between a pair of vertices is the  number of multiple connections they have \cite{Newman2004,Serrano2005}. Now, the expected quantity of a given vertex is no longer its degree but its strength. Therefore, let us change notation and denote by $P(s)\sim s^{-(1+\tau)}$, $0<\tau \le 1$, the distribution of expected strengths, that is, the number of expected connections of vertices. 

At the level of strength, vertices are uncorrelated and the average weight of edges linking nodes of strengths $s_i$ and $s_j$ can be written as
\begin{equation}
\bar{\omega}_{ij}=\frac{s_is_j}{2S},
\label{eq:5}
\end{equation}
where $S$ is the total number of connections --regardless they are multiple or not-- in the network, that is, 
\begin{equation}
S= \frac{1}{2}\sum_i s_i\sim N^{1/\tau}
\end{equation}
The average degree of a vertex can be obtained as
\begin{equation}
\bar{k}_i=\sum_j \bar{\omega}_{ij} \Theta(1-\bar{\omega}_{ij})+\sum_j  \Theta(\bar{\omega}_{ij}-1),
\end{equation}
where $ \Theta(\cdot)$ is the Heaviside step function. The first term in this equation is the contribution of those connections that, on average, are smaller than 1. The second term represents connections with an average number larger than 1 and, thus, holding multiple connections. In this case, the contribution to the degree is just 1 and not $\bar{\omega}_{ij}$. Using the expression for the weights, Eq.~(\ref{eq:5}), in the continuum limit we can write
\begin{equation}
\bar{k}(s_i)=N\int_1^{2S/s_i} P(s_j) \frac{s_is_j}{2S}ds_j+N\int_{2S/s_i}^{\infty} P(s_j)ds_j
\end{equation}
that, in the thermodynamic limit goes as
\begin{equation}
\bar{k}(s_i)\sim \frac{1}{\tau^{\tau}(1-\tau)^{1-\tau}}s_i^{\tau}.
\end{equation}
It is worth noticing that there are finite size effects that depend on the specific form of the strength distribution. The model generates a non trivial weighted network with an exponent $\beta=1/\tau$ larger than 1, as it is observed in real weighted networks \cite{Barrat04a}.

The tail of the degree distribution can be obtained from this scaling relation, yielding an exponent $\gamma=2$. Thus, even though the expected strength distribution has an exponent smaller than two, the resulting network has exponent equal to two and a number of connections growing as $\sum_i k_i \sim N\ln N$.

\begin{figure}
  \includegraphics[height=.4\textheight]{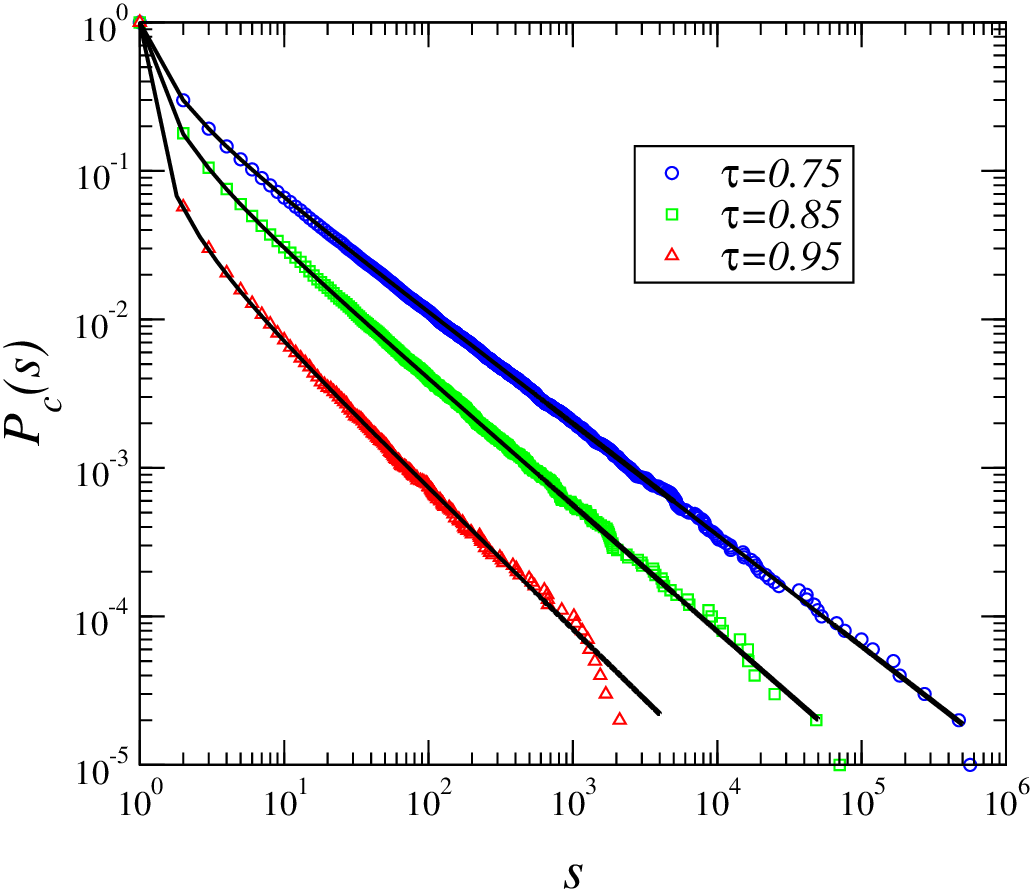}
  \caption{Cumulative strength distributions, that is, $P_c(s)=\sum_{s'\ge s} P(s')$, used in the simulations. The solid lines correspond to the theoretical curves given by Eq.~(\ref{eq:11}) }
\label{fig1}
\end{figure}

\section{Numerical Simulations}

\begin{figure}
  \includegraphics[height=.4\textheight]{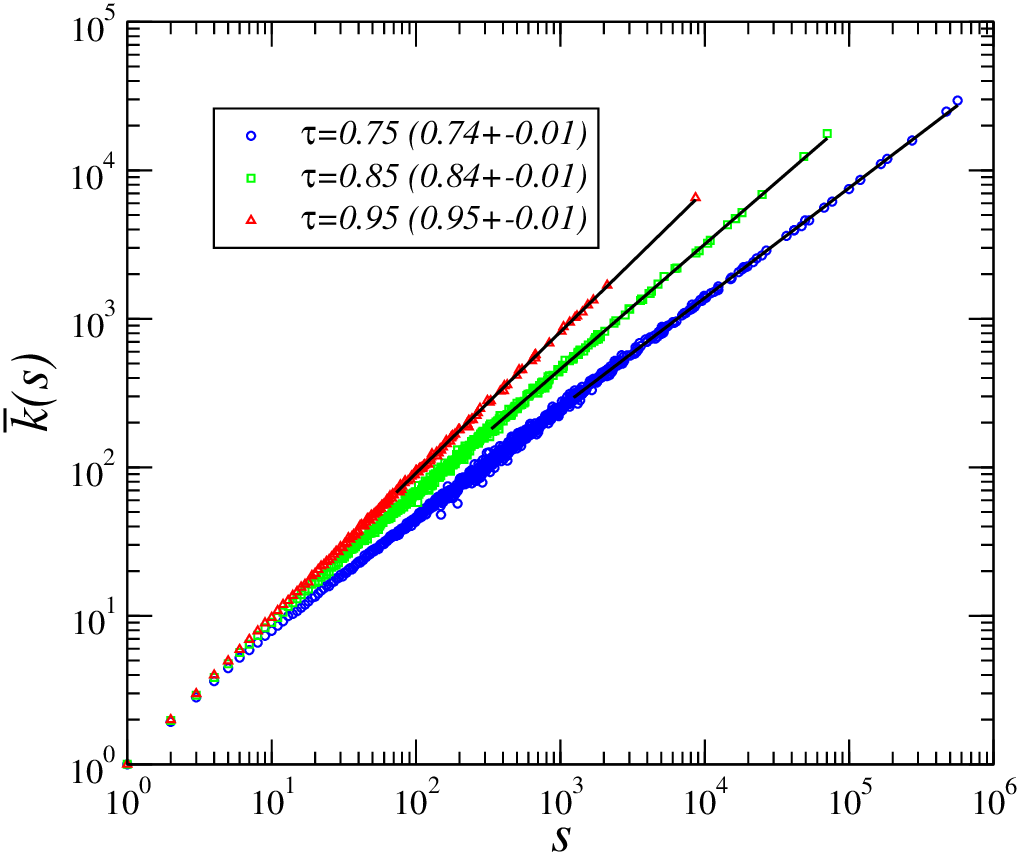}
  \caption{Simulation results for the average degree of vertices of strength $s$, $\bar{k}(s)$, as a function of $s$, for $\tau=0.95$, $\tau=0.85$, and $\tau=0.75$. The size of the network is $N=10^5$. In all cases, the strength distribution is given by Eq.~(\ref{eq:10}) with $s_c=\tau$. Solid lines correspond to the best fit estimates (numbers in parenthesis into the legend box) of the theoretical behavior $\bar{k}(s) \sim s^{\tau}$.}
  \label{fig2}
\end{figure}

To test the results of our analysis, we have performed numerical simulations of the configuration model, generating expected strength distributions of the form
\begin{equation}
P(s)\sim \frac{1}{(s-s_c)^{1+\tau}}.
\label{eq:10}
\end{equation}
The value of $s_c$ modulates the probability to find strengths of value larger than $1$. We choose this particular form because, by an appropriate choice of $s_c$, the finite size effects are minimized. In all the simulations, the size of the network is $N=10^5$, the values of $\tau$ are $1.95$, $1.85$, and $1.75$ and $s_c=\tau$. We first show, in Fig.~\ref{fig1}, the cumulative strength distributions used in the simulations as compared to the theoretical ones,
\begin{equation}
P_c(s)=\frac{(1-\tau)^{\tau}}{(s-\tau)^{\tau}},
\label{eq:11}
\end{equation}
computed using the continuum approximation. 

\begin{figure}
  \includegraphics[height=.4\textheight]{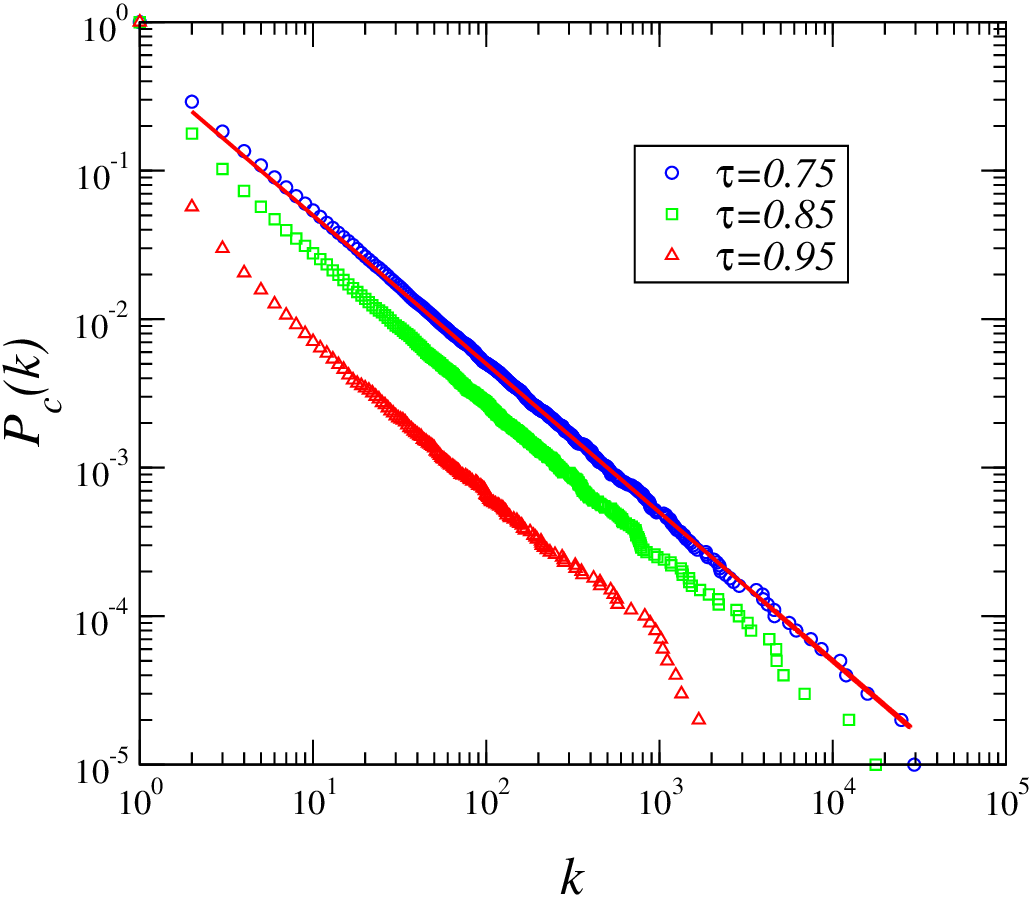}
  \caption{Cumulative degree distributions, $P_c(k)=\sum_{k'=k} P(k')$, generated by de model. The solid line is power law of the form $k^{-1}$ corresponding to an exponent $\gamma=2$.}
\label{fig3}
\end{figure}

Fig.~\ref{fig2} shows the scaling relation between $\bar{k}(s)$ and the strength $s$. As it can be seen, the theoretical prediction $s^{\tau}$ is well satisfied, with the following estimates for the exponent $\beta^{-1}$:  $\beta^{-1}=0.74 \pm 0.01$ for $\tau=0.75$, $\beta^{-1}=0.84 \pm 0.01$ for $\tau=0.85$, and $\beta^{-1}=0.95 \pm 0.01$ for $\tau=0.95$. Finally, the resulting cumulative degree distribution is plotted in Fig.~\ref{fig3}, showing that this function goes, for large degrees, as $k^{-1}$ independent of $\tau$, as predicted by our analysis.

\section{Conclusions}

In summary, we have analyzed the behavior of the classical configuration model in the case of expected strength distributions following a power law form of exponent smaller than two. We have shown that, in this case, the resulting network is weighted, where the weight stands for the number of multiple connections among vertices. The model presents a non trivial scaling relation between strength and degree and a degree distribution with exponent $\gamma=2$, independent of the exponent of the strength distribution, $\tau$. These results highlight the subtleties that may arise when dealing with SF networks even in the most simplified models.

\begin{theacknowledgments}
This work has been partially supported by DGES of the Spanish government,
Grant No. FIS2004-05923-CO2-02, and EC-FET Open project COSIN
IST-2001-33555. M. B.  acknowledges financial support from the
MCyT (Spain).
\end{theacknowledgments}

\end{document}